\documentclass[a4paper,11pt]{article}
\usepackage{aaskaiid}
\usepackage{orcidlink}

\newcommand\aastar{AA$^{*}$}

\usepackage[separate-uncertainty=true,multi-part-units=single]{siunitx}
\DeclareSIUnit\jansky{Jy}
\DeclareSIUnit\parsec{pc}
\DeclareSIUnit\erg{erg}

\title{The Variability of Radio Stars}
\ShortTitle{The variability of radio stars}

\author[1]{Laura N. Driessen\orcidlink{0000-0002-4405-3273}}
\ShortName{Driessen et al.} 
\author[2]{Alex Andersson\orcidlink{0000-0003-2734-1895}}
\author[3,4]{Joseph R. Callingham\orcidlink{0000-0002-7167-1819}}
\author[5,6]{Barnali Das\orcidlink{0000-0001-8704-1822}}
\author[4]{Jakob van den Eijnden\orcidlink{0000-0002-5686-0611}}
\author[1,7]{Tara Murphy\orcidlink{0000-0002-2686-438X}}
\author[1,8]{Kovi Rose\orcidlink{0000-0002-7329-3209}}
\author[8]{Joshua Pritchard\orcidlink{0000-0003-1575-5249}}
\author[9]{Miguel P\'erez-Torres\orcidlink{0000-0001-5654-0266}}
\author[]{the Transients Science Working Group}

\affiliation[1]{Sydney Institute for Astronomy, School of Physics, The University of Sydney, New South Wales 2006, Australia}
\emailAdd{laura.driessen@sydney.edu.au}
\affiliation[2]{Astrophysics, Department of Physics, University of Oxford, Keble Road, Oxford OX1 3RH, UK}
\affiliation[3]{ASTRON, Netherlands Institute for Radio Astronomy, Oude Hoogeveensedijk 4, Dwingeloo, 7991\,PD, The Netherlands}
\affiliation[4]{Anton Pannekoek Institute for Astronomy, Universiteit van Amsterdam, Science Park 904, 1098, XH, Amsterdam, The Netherlands}
\affiliation[5]{CSIRO, Space and Astronomy, P.O. Box 1130, Bentley, WA 6102, Australia}
\affiliation[6]{National Centre for Radio Astrophysics, Tata Institute of Fundamental Research, Pune University Campus, Pune-411007, India}
\affiliation[7]{ARC Centre of Excellence for Gravitational Wave Discovery (OzGrav), Australia}
\emailAdd{tara.murphy@sydney.edu.au}
\affiliation[8]{Australia Telescope National Facility, CSIRO, Space \& Astronomy, PO Box 76, Epping, NSW 1710, Australia}
\affiliation[9]{Instituto de Astrof\'isica de Andaluc\'ia (IAA-CSIC), Glorieta de la Astronom\'ia s/n, E-18008 Granada, Spain
}

\abstract{Stellar radio emission is highly variable with stellar flares lasting from milliseconds to hours. For some stars, their flares or bursts can repeat once every couple of hours, while other stars may flare only once in hundreds of hours of observing. Some stars and stellar systems vary slowly over months to years. In this chapter we present the current methods for identifying radio stars and the important role that variability plays in these detection methods. We highlight that radio stars are the second most common radio variable object in image-plane searches for radio variable sources. We also present our predictions for the number of stars both the SKA-mid and SKA-low \aastar\ arrays could detect based on searches using SKA pathfinders and precursors.
}

\begin{document}
\newcommand{\actaa}{Acta Astron.} 
\newcommand{\araa}{ARA\&A} 
\newcommand{\aar}{A\&ARv} 
\newcommand{\aapr}{A\&ARv} 
\newcommand{\ab}{Astrobiol.} 
\newcommand{\aj}{AJ} 
\newcommand{\apj}{ApJ} 
\newcommand{\apjl}{ApJL} 
\newcommand{\apjs}{ApJSS} 
\newcommand{\ao}{Appl. Opt.} 
\newcommand{\apss}{Astro. \& Space Sci.} 
\newcommand{\aap}{A\&A} 
\newcommand{\aaps}{A\&AS.} 
\newcommand{\baas}{Bull. Am. Astron. Soc.} 
\newcommand{\caa}{Chinese A\&A} 
\newcommand{\cjaa}{Chinese J. A\&A} 
\newcommand{\cqg}{Class. Quantum Gravity} 
\newcommand{\gal}{Galaxies} 
\newcommand{\gca}{Geo. Cosmo. Acta} 
\newcommand{\icarus}{Icarus} 
\newcommand{\jcap}{JCAP} 
\newcommand{\jgr}{J. Geophys. Res.} 
\newcommand{\jgrp}{J. Geophys. Res. Planets} 
\newcommand{\jqsrt}{J. Quant. Spectrosc. Radiat. Transf.} 
\newcommand{\memsai}{Mem. SAIt} 
\newcommand{\mnras}{MNRAS} 
\newcommand{\nat}{Nature} 
\newcommand{\nastro}{Nat. Astron.} 
\newcommand{\ncomms}{Nat. Commun.} 
\newcommand{\nphys}{Nat. Phys.} 
\newcommand{\na}{New Astron.} 
\newcommand{\nar}{New Astron. Rev.} 
\newcommand{\physrep}{Phys. Rep.} 
\newcommand{\pra}{Phys. Rev. A} 
\newcommand{\prb}{Phys. Rev. B} 
\newcommand{\prc}{Phys. Rev. C} 
\newcommand{\prd}{Phys. Rev. D} 
\newcommand{\pre}{Phys. Rev. E} 
\newcommand{\prx}{Phys. Rev. X} 
\newcommand{\prl}{Phys. Rev. Let.} 
\newcommand{\psj}{Planet. Sci. J.} 
\newcommand{\planss}{Planet. Space Sci.} 
\newcommand{\pnas}{Proc. Natl Acad. Sci. USA} 
\newcommand{\procspie}{Proc. SPIE} 
\newcommand{\pasa}{PASA} 
\newcommand{\pasj}{PASJ} 
\newcommand{\pasp}{PASP} 
\newcommand{\rmxaa}{RMXAA} 
\newcommand{\sci}{Science} 
\newcommand{\sciadv}{Sci. Adv.} 
\newcommand{\solphys}{Sol. Phys.} 
\newcommand{\sovast}{Soviet Ast.} 
\newcommand{\ssr}{Space Sci. Rev.} 
\newcommand{\uni}{Universe} 

\maketitle

\section{Introduction}

\begin{figure}
\includegraphics[width=\columnwidth]{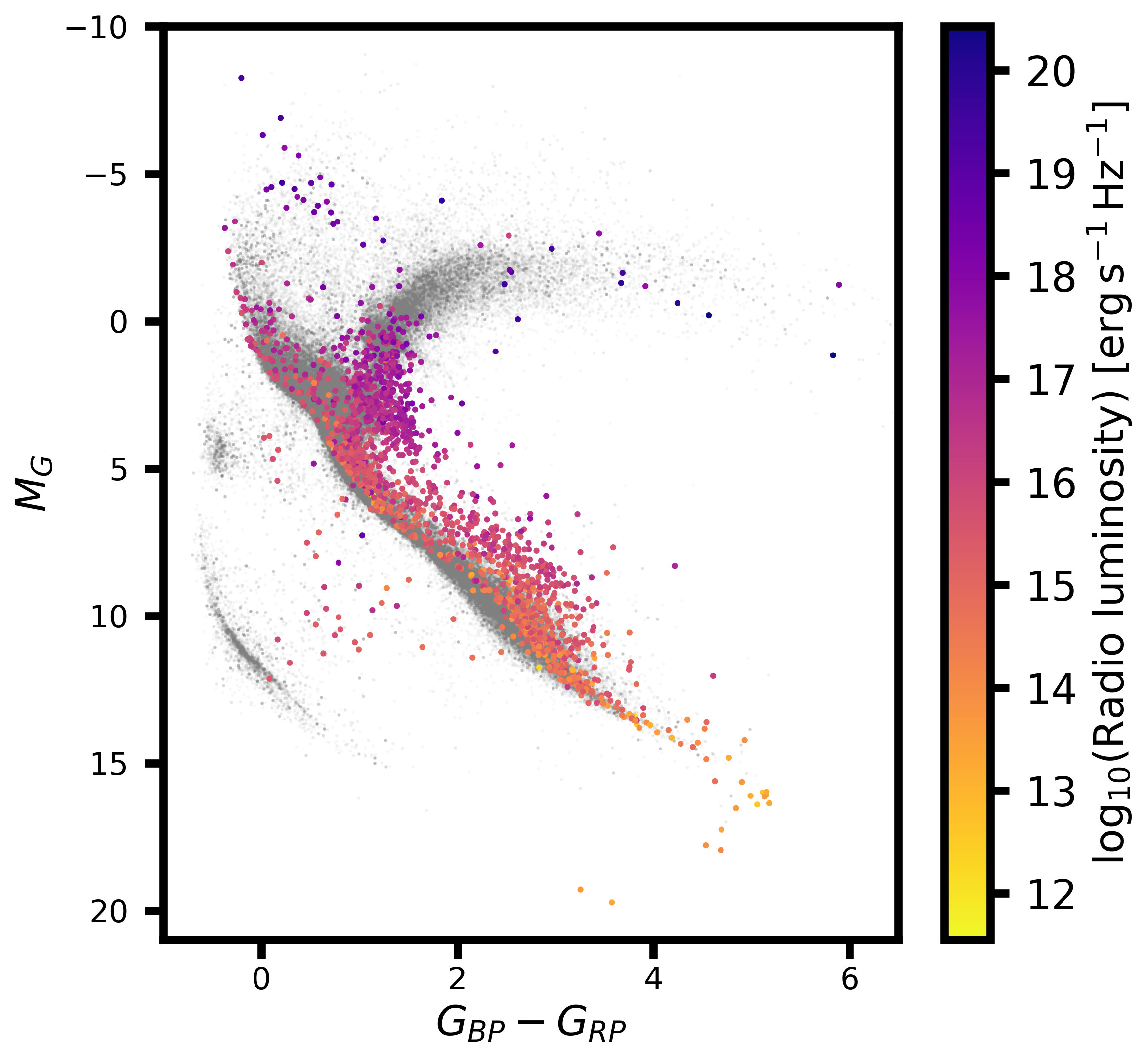}
\caption{Colour-magnitude diagram of radio stars in version 1.1 of the Sydney Radio Star Catalogue \citep{SRSC}. This diagram includes more than 1,500 radio detected stars. The radio luminosity for each star is calculated using the minimum detected flux density and the median \textsc{rgeo} distance from \citet{2021AJ....161..147B}.
}
\label{fig: CMD}
\end{figure}

Stars spanning the entire Hertzsprung-Russel diagram (Fig.\,\ref{fig: CMD}) have been observed to emit radio waves \citep[e.g.][]{2019PASP..131a6001M,2025arXiv250921467M,2002ARA&A..40..217G,SRSC}. This includes the coolest stars, ultra-cool dwarfs (UCDs) down to T8 \citep[e.g.][]{Rose_BD}, main-sequence stars \citep{2019PASP..131a6001M}, and some of the most extreme stellar systems, colliding wind binaries \citep[e.g.][]{1976ApJ...203L..35S,1986ApJ...303..239A,2005ApJ...623..447D}. We can use radio detections of stars to investigate their coronae, magnetic fields \citep[][]{1985ARA&A..23..169D,2002ARA&A..40..217G,2019PASP..131a6001M}, and possibly interaction between stars and their planets \citep[e.g.][]{2020NatAs...4..577V,2024MNRAS.528.2136S}.

Radio emission from stars can be generated via both incoherent and coherent mechanisms. The key emission mechanisms to consider for radio stars are: gyrosynchrotron emission, electron-cyclotron maser (ECM) emission, and plasma emission. Gyrosynchrotron emission is incoherent and can be observed as either persistent, variable, or flaring radio emission and can be unpolarised or circularly polarised. Both ECM and plasma emission are coherent and are typically observed as highly circularly polarised flares on time scales of minutes to hours. 
Massive stars are known to emit thermal free-free emission with strongly inverted spectra \citep{wright1975,Panagia1975}. The radio emission from these stars is relatively faint compared to their luminosities at other wavelengths. We will not discuss thermal emission from radio stars in detail in this chapter.
See \citet{2002ARA&A..40..217G} and \citet{1985ARA&A..23..169D} for reviews of stellar radio emission mechanisms.

\begin{figure}
\includegraphics[width=\columnwidth]{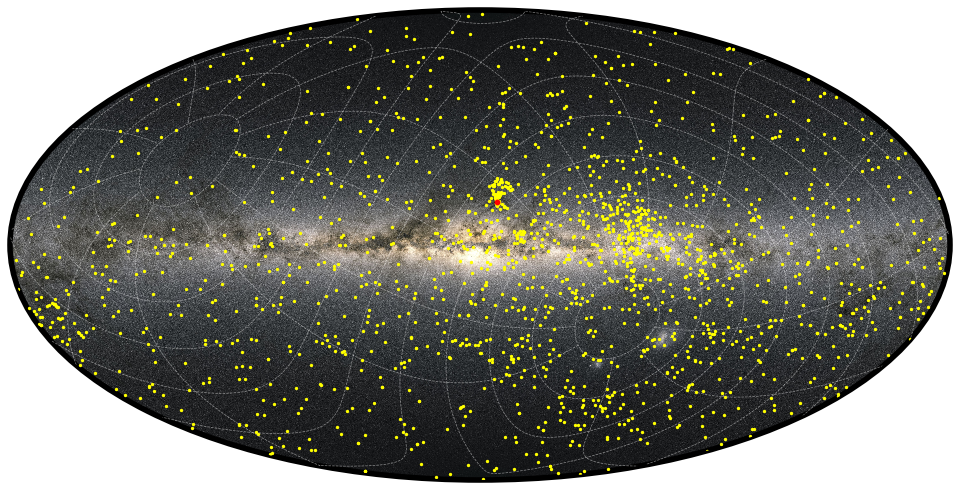}
\caption{Sky map of the stars, shown in galactic coordinates with the equatorial coordinate frame as grey-dashed lines, in version 1.1 of the Sydney Radio Star Catalogue \citep[SRSCv1.1;][]{SRSC}.The northern hemisphere is not well sampled as the SRSC includes star mainly identified using Australian SKA Pathfinder \citep[ASKAP;][]{2021PASA...38....9H} observations. The small cluster of radio stars slightly above the Galactic Centre and marked with a red dot are members of the Rho Ophiuchi cloud complex.
}
\label{fig: SRSC sky map}
\end{figure}

In wide-field radio surveys, radio stars appear as unresolved radio sources that can be observed as unpolarised \citep[e.g.][]{2017A&A...598A..42H} or highly circularly \citep[e.g.][]{2006ApJ...637.1016O} or elliptically polarised \citep[e.g.][]{Zic_UVCeti}, persistent or variable on time scales from milliseconds \citep{ADLeo_FAST} to hours \citep[e.g.][]{Villadsen22}, flaring \citep[e.g.][]{Andersson_Star} or periodic \citep[e.g.][]{2025arXiv250703882D}. Radio stars are detected from tens of MHz \citep[e.g.][]{2025arXiv250607912Z} up to hundreds of GHz \citep[e.g.][]{2025ApJ...982...43B}.
Due to their typical radio luminosities and current telescope sensitivity limits, the radio stars detected to date have typically been within 1\,kpc and are therefore isotropically distributed on the sky, see Fig. \,\ref{fig: SRSC sky map}. 

Identifying and investigating radio stars and sub-stellar objects is essential for understanding particle acceleration and magnetospheric currents in stellar atmospheres \citep{Villadsen22}, probing space weather environments \citep{Zic_proxima}, connecting our understanding of stellar magnetic fields to planetary magnetic fields \citep{Kavanagh2024} and potentially investigating star-planet interactions \citep{2024NatAs...8.1359C}. For massive early-type stars, often residing in binary systems, radio observations can probe the properties of their stellar winds, the non-thermal emission from interactions within the binary, and its feedback on its surroundings. 

So far only a tiny fraction of stars have been detected in radio.
The \textit{Gaia} Data Release 3 \citep[DR3;][]{GaiaDR3} contains the full astrometric solutions for nearly 1.5 billion sources. This amounts to approximately 2\% of the Milky Way. Prior to SRSCv1.1, we had identified fewer than 1500 radio stars \citep{SRSC}. The current SKA pathfinders and precursors have already significantly increased the known number of radio stars. This is compared to the previous, canonical radio star catalogue ``The Catalogue of Radio Stars'' \citep{Wendker_1978,Wendker_1987,Wendker_1995}, which was most recently updated in 2001. The sensitivity, time resolution, and survey strategies of the SKA telescopes will dramatically increase these numbers.

A key consideration for any search for or targeted observation of stellar radio emission is variability \citep[see e.g.][for reviews on flaring time scales]{1990SoPh..130..265B,2002ARA&A..40..217G}. This radio variability means that in a single epoch of a survey you see one set of radio stars, and in a second epoch see an almost independent set of radio stars. This impacts all of the search methods presented in Section\,\ref{sec: identifying radio stars}. It also means that radio stars are commonly found in untargeted searches for variable and transient radio sources. In fact, radio stars are the second most common variable source identified in untargeted, image-plane searches for radio variables behind Active Galactic Nucleus (AGN) variability \citep{2026arXiv260222739D} and are the most common source identified in searches for long-period radio transients \citep{2026MNRAS.545f2008L}. This means that is it essential to consider variability when searching specifically for stellar radio emission, and it is essential to consider stellar radio emission when searching for variable radio sources.

Radio investigations of stars and sub-stellar objects are presented in other chapters of this book \citep{Cavallaro01.2026.SKA,Kavanagh01.2026.SKA,Vedantham01.2026.SKA}. In this chapter we will explore the observed properties of radio stars, how we identify radio stars, and SKAO survey rate expectations and implications for SKAO variability searches.

\section{Identifying radio stars}
\label{sec: identifying radio stars}

Identifying stars in wide-field radio surveys is challenging as it is difficult to distinguish stellar radio emission from AGN based on continuum image detections alone. 
In radio images, the most common compact source class by far is AGN \citep{2015ApJ...801...26H}. In Stokes I images, even with frequency spectral information, there is no signature to distinguish radio emission from a background AGN compared to radio emission from a foreground star. Cross-matching must be performed with care due to the high chance-coincidence rate between radio AGN and optical stars \citep[][]{2019RNAAS...3...37C}.

Traditionally, most radio observations of stars have been targeted observations of objects known to be active in other wavebands \citep[e.g.][]{2018ApJ...862..113C,PerezTorres2021,2024A&A...682A.170B}.
Recently, methods for identifying radio stars have been developed and refined to search for stars in the large data sets from SKA pathfinder and precursor instruments. These include untargeted variability searches, cross-matching, and circular polarisation searches. We briefly summarise these approaches in the following sections.

\subsection{Targeted follow-up}

Historically, radio stars have been identified by targeting stars with other activity indicators such as H$\alpha$ emission, Calcium H and K emission, rotation rate, optical flares and X-ray activity. This is because these are indicators of stellar activity and are therefore indicators of stellar magnetic fields \citep{2019LNP...955.....L}. Targeting these stars, or stars that are already known radio stars, increases the probability of observing stellar radio emission. While this approach is fruitful \citep[e.g.][]{Villadsen22}, it introduces significant biases in the detected population, effectively excluding the possibility of detecting radio emission from less-active stellar systems. The recent increases in field of view, survey speed, and sensitivity of radio telescopes mean that we can use other methods to identify radio stars. This method remains essential for gathering the long integration observations required for dynamic spectra to study stellar emission mechanisms in detail.

\subsection{Variability}

Radio stars can be highly variable \citep[e.g.][]{1990SoPh..130..265B,2002ARA&A..40..217G} and can vary on time scales of milliseconds to hours or even years.
This variability can be caused by stochastic flares, highly beamed radio emission, or variable persistent emission. All of these can be modulated by the rotation period of a star, the orbital period of a binary system, or a combination of these.

Thermal radio emission from massive single and binary stars is also expected to vary, though on longer time scales -- days to months -- compared to non-thermal stellar radio emission. Evidence for such variability in thermal wind emission has been seen with ALMA, with variability amplitudes of a factor $\sim 2$, but not yet at radio frequencies \citep{vandeneijnden2025}. Massive stars in interacting binary systems, such as Colliding Wind Binaries and $\gamma$-ray binaries, instead show variable and bright non-thermal radio emission that has routinely been studied by SKA pathfinders. 

\begin{figure}
\includegraphics[width=\columnwidth]{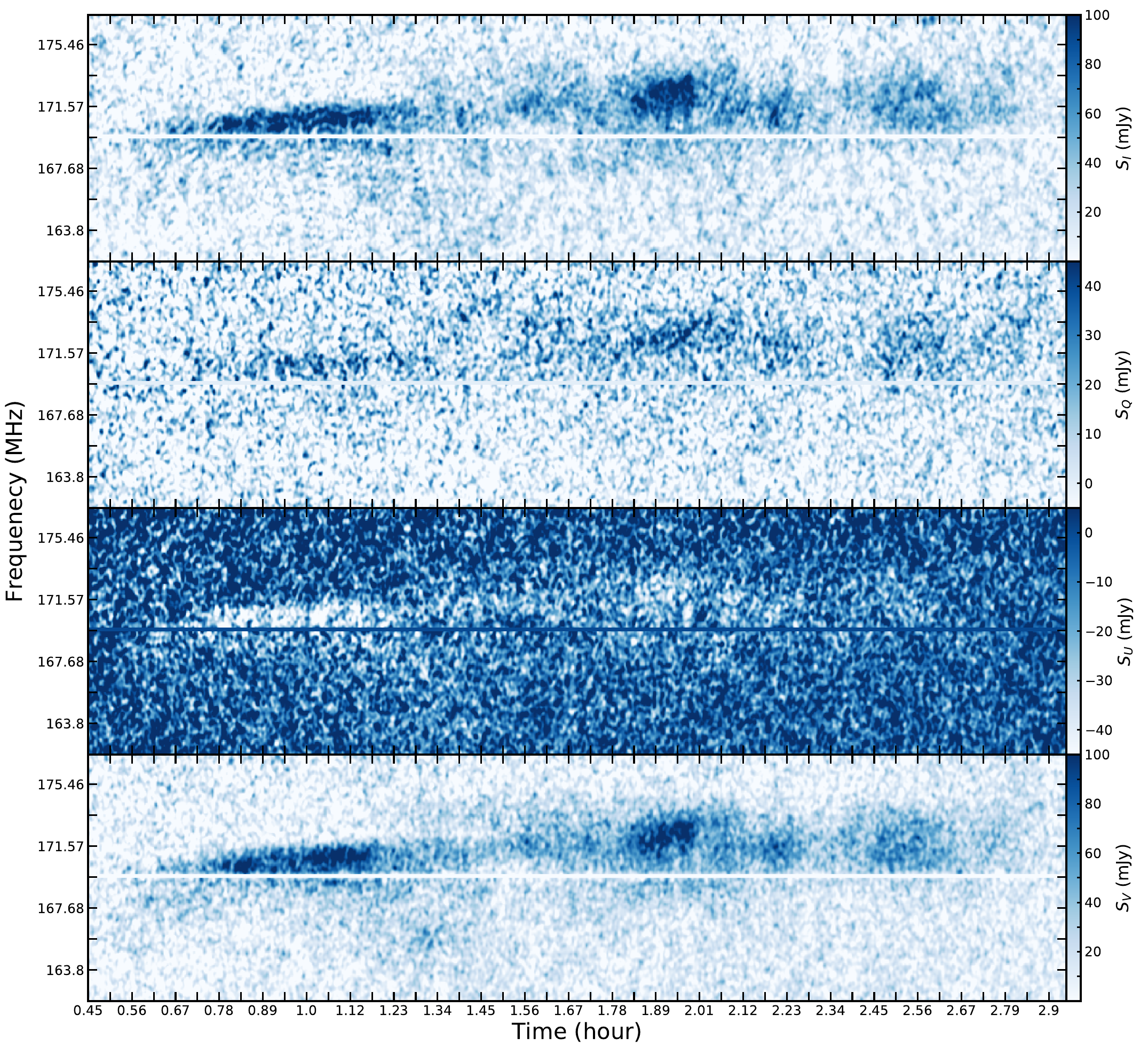}
\caption{LOFAR dynamic spectrum of star CR Draconis \citep{2021A&A...648A..13C}. The panels from top to bottom show the Stokes I, Q, U and V dynamic spectra.
}
\label{fig: lofar ds}
\end{figure}

\begin{figure}
\includegraphics[width=1.0\textwidth]{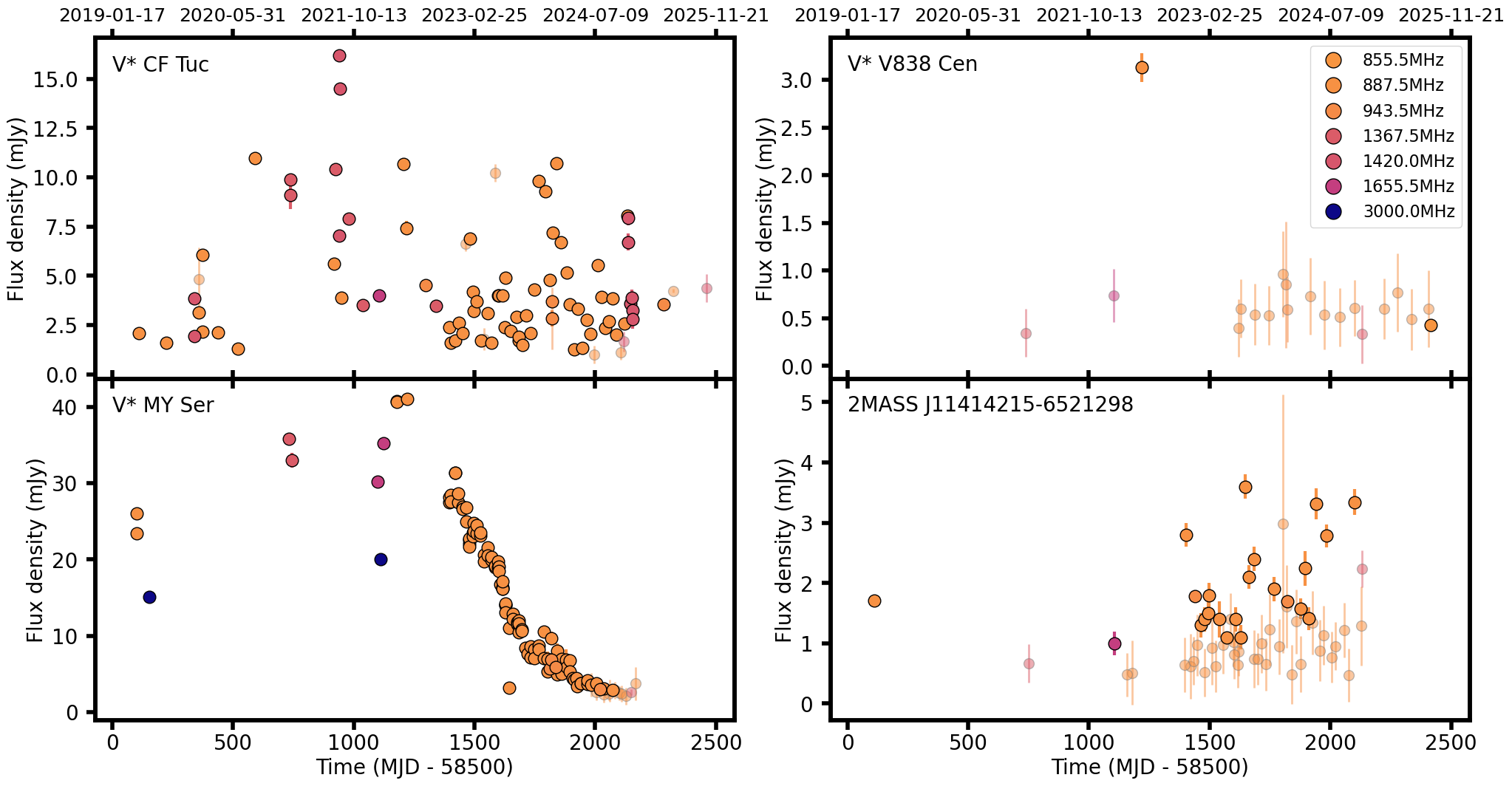}
\caption{Light curves of four radio stars using ASKAP data. The solid points are $5\sigma$ detections from the automated {\sc selavy} \citep{2012MNRAS.421.3242W,2012PASA...29..371W} source extraction pipeline. The transparent points are aperture fitting of non-detections in ASKAP images covering the location of the star. CF Tuc is an RS CVn binary, V383 Cen is a BY Draconis type star. MY Ser is a colliding wind binary. 2MASS J11414215$-$6521298 is a candidate young stellar object.
}
\label{fig: light curves}
\end{figure}

We can see in Fig.\,\ref{fig: lofar ds} an example of a dynamic spectrum of a star varying on time scales shorter than an hour. Dynamic spectra are essential for identifying and investigating stellar radio emission on time scales of seconds to hours. Fig.\ref{fig: light curves} shows light curves of four stars using ASKAP data. These light curves demonstrate that, in imaging data, radio stars can vary from epoch to epoch or over years. Fig.\,\ref{fig: light curves} also shows the diversity in flux density and light curve shapes.

The variability of stellar radio emission presents both a challenge and an opportunity. In targeted follow-up, stellar variability could mean that no radio emission is observed in a given observation despite the star being a known radio star. For both cross-matching and circular polarisation searches, cross-matching to a single radio epoch would mean a large fraction of stars would be missing due to being ``off'' in that single epoch. This means that multi-epoch observations or long integration observations that can be divided into shorter observation images or dynamic spectra are essential for identifying and investigating radio stars.

Stellar variability presents the opportunity to identify many new radio stars in untargeted searches for radio variability. It is emerging in current image-plane radio searches for variability that radio stars are the most commonly identified radio variable sources behind AGN \citep[either AGN scintillation or intrinsic variability;][]{2026arXiv260222739D}. In searches designed to identify new long-period radio transients using short time scale imaging and dynamic spectra, radio stars are emerging as the most common radio variable source identified \citep{2026MNRAS.545f2008L}.

\subsection{Cross-matching}
\label{sec: circ pol}

Cross-matching is a standard way of associating sources from two different catalogues based on their positions \citep[e.g.][]{2015ApJ...801...26H,2021PASA...38...41F}. For example, matching stars from a optical survey, e.g. Hipparcos, to unresolved sources in a radio survey, e.g. FIRST \citep{1999AJ....117.1568H}. Sources where the coordinates of the star match the coordinates of the radio source within some radius are considered matches.

Cross-matching is straightforward to implement. However, a key challenge with this method is that most optical point sources are stars but most radio point sources are AGN \citep[e.g.][]{2015ApJ...801...26H}. This means that a direct cross-match between e.g. \textit{Gaia} DR3 and VLASS would result in mostly spurious matches between a star and an AGN, due to chance positional alignment or ``chance coincidence'' \citep{2019RNAAS...3...37C}. With only the information available in each survey, there is no way to identify which cross-match is a chance match and which is a true match.

This method can be successful if care is taken to select both the surveys used for cross-matching and the match radii. This can be done using Monte Carlo simulations \citep[e.g.][]{SRSC}.

Multi-epoch observations are important for this search method as many stars may be missed in cross-matching if they are not flaring during the observation.

\subsection{Circular polarisation}

Stellar radio emission mechanisms, gyrosynchrotron, ECM and plasma emission,  result in high circular or elliptical polarisation fractions \citep{1985ARA&A..23..169D}. Radio emission from stars can have circular polarisation fractions up to 100\%. \citet{2021MNRAS.502.5438P} and \citet{vlotss} demonstrated that, in circular polarisation searches, radio stars have typical polarisation fractions of 20--100\% while pulsar have typical polarisation fractions of 10--30\%.

High circular polarisation fractions ($\gtrsim10\%$) from stellar radio emission can also be used to exclude AGN from cross-matching. This is because radio AGN have circular polarisation fractions $\lesssim3\%$ \citep[e.g.][]{1988ARA&A..26...93S,2000ApJ...545..798M,2013MNRAS.435..311O}. This makes cross-matching optical surveys with circularly polarised radio sources more reliable for identifying stars than cross-matching with full polarisation sources.

Cross-matching highly circularly polarised radio sources to optical surveys has been used to successfully identify handfuls of new radio stars. This method has been performed using the Low Frequency Array \citep[LOFAR;][]{2013A&A...556A...2V} and ASKAP. For example, \citet{2021MNRAS.502.5438P} identified 23 new radio stars, \citet{2024MNRAS.tmp..161P} identified 36 radio stars, \citet{2021NatAs...5.1233C} and \citet{2021A&A...654A..21T} identified 18 new radio stars, and \citet{vlotss} included the 36 stars identified in LOFAR data using circular polarisation searches.

\begin{figure}[!ht]
\centering
\includegraphics[width=9.0 cm]{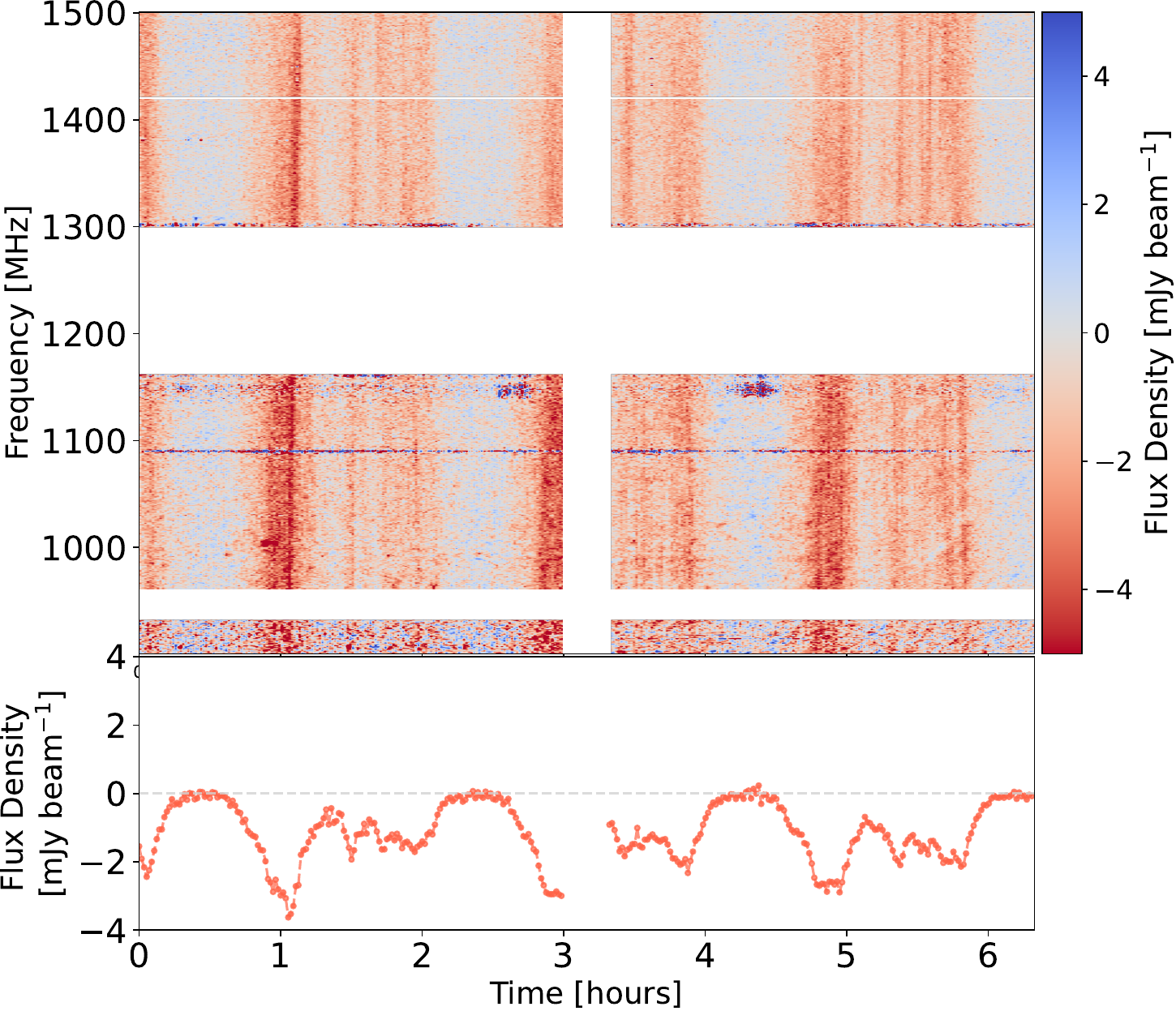}
\includegraphics[width=6.0 cm]{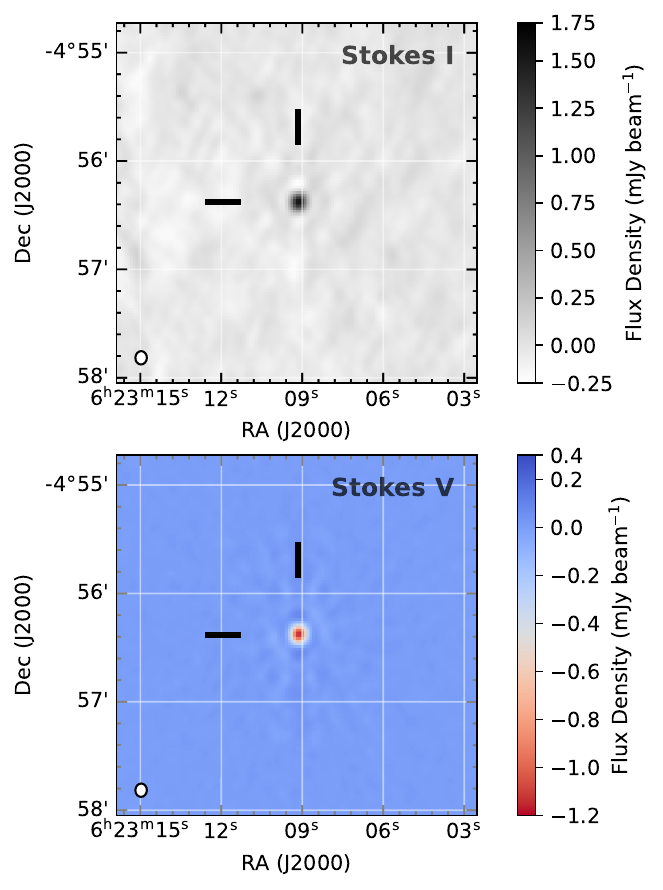}
\caption{\textit{Left:} Stokes V dynamic spectrum ($0.9$--$1.5$\,GHz) of a T8 brown dwarf observed with MeerKAT; adapted from \cite{Rose_BD}. The Stokes V lightcurve (lower panel) and the dynamic spectrum display periodic behaviour with complex sub-pulse structure.
\textit{Right:} Stokes I and Stokes V continuum detection images from the same observation.}
\label{fig: Rose BD}
\end{figure}

We can see an example of a highly circularly polarised radio emitting brown dwarf in Fig. \,\ref{fig: Rose BD}. This is a sub-stellar source identified using ASKAP Stokes V images. It also demonstrates how important both Stokes V images and dynamic spectra are for finding and analysing radio stars.

While circular polarisation searches have been highly successful using SKA pathfinders and precursors, it is important to note that not all stellar radio emission is strongly circularly polarised.

Circularly polarised stellar emission is variable by nature. Stellar flares can last up to a few hours, but some stars are known to flare only once every $\sim$\SI{70}{\hour} (Mitchell-Bolton, J., in prep.). Both of these factors mean that variability and multi-epoch observations are key to finding new radio stars and studying the polarisation properties of known radio stars.

\subsection{Proper motion}

Proper motion has been used in the past to distinguish between stellar radio emission and radio-bright AGN. This has typically been done using Very Long Baseline Interferometry (VLBI) \citep[e.g.][]{1992A&A...258..112L}. More recently, \citet{Driessen_ProperMotion} used observations over thirty years apart to identify eight radio stars, including two that  had not been previously identified. While this method has only been used to identify a small number of radio stars, multi-epoch surveys such as the Rapid ASKAP Continuum Surveys \citep[RACS;][]{2020PASA...37...48M,2021PASA...38...58H,2023PASA...40...34D} and higher resolution surveys with the SKA telescopes will greatly increase the sample size. There will be more than forty years between the initial Karl G. Jansky Very Large Array \citep[VLA;][]{perley2011} Faint Images of the Radio Sky at Twenty-centimetres \citep[FIRST;][performed between 1993 and 2011]{1994ASPC...61..165B,1995ApJ...450..559B} survey observations and the initial SKA-mid telescope surveys, and over ten years between the first RACS surveys and SKA-mid telescope surveys. The longer time between observations, repeat sky coverage by RACS, and higher resolution of SKA-mid means that the motions of stars with lower proper motions can be identified.

\section{Population statistics and prospects for the SKA}

Fewer than 1500 stars are know to emit in the radio. However, we can use the recent searches using SKA pathfinder and precursor instruments to investigate the population and flare statistics. We can use these statistics to predict the detection rates for the SKA \aastar\ arrays, both mid and low.

\subsection{Mid frequencies}

Circular polarisation searches of the RACS-low and VAST pilot surveys with
ASKAP \citep{2021MNRAS.502.5438P, 2024MNRAS.tmp..161P} produced a total of 229
detections of a sample of 76 stars, with the majority of detections
attributed to coherent bursts from M-dwarfs. These detections
correspond to instantaneous surface densities of
$7.1_{-1.4}^{+1.6} \times 10^{-4}$ \si{\deg^{-2}} and \SI{1.7 \pm
  0.2 e-3}{\deg^{-2}} for M-dwarfs and the wider radio star
population respectively, when sampled on \SI{12}{\minute} time scales.

The radio luminosity distributions of this sample peak at
$\sim$\SI{e15}{\erg\per\second\per\hertz} for M-dwarfs and young
stellar objects, and $\sim$\SI{e17}{\erg\per\second\per\hertz} for
interacting, chromospherically active binary systems and hot magnetic
stars. However, the sensitivity limits of these surveys prohibit
detection of systems below $\sim$\SI{e14}{\erg\per\second\per\hertz}
beyond a distance of \SI{10}{\parsec} and
$\sim$\SI{e15}{\erg\per\second\per\hertz} beyond \SI{25}{\parsec}.

Through multi-epoch sampling of the VAST pilot survey
\citet{2024MNRAS.tmp..161P} provided constraints on the activity fraction of
the M-dwarf population, determining that at least \SI{10 \pm
  3}{\percent} of M-dwarfs should produce coherent radio bursts. Most
M-dwarfs were only detected once, producing an upper limit of
\SI{8.5}{\percent} on the population mean detection fraction. As each
detection was attributed to a coherent burst, this is also equivalent to
an estimate of the burst duty cycle, and allows a lower bound estimate
of $9_{-7}^{+11} \times 10^{-3}$  \si{\deg^{-2}} M-dwarfs capable of
producing coherent bursts above \SI{e15}{\erg\per\second\per\hertz}.

Assuming the current projected SKA-Mid \aastar\ \SI{1}{\hour} integration
continuum sensitivity of $\sim$\SI{20}{\micro\jansky} and
$\sim$\SI{10}{\micro\jansky} in Band 1 and Band 2 respectively,
\SI{1}{\hour} pointings with SKA-Mid will push the 5$\sigma$
\SI{e15}{\erg\per\second\per\hertz} completeness horizon out to
$\sim${100}\si{\parsec} and probe the luminosity distribution down to
\SI{e13}{\erg\per\second\per\hertz} for systems within
$\sim$\SI{10}{\parsec}. Within these distances the distribution of
stellar sources should be independent of direction, allowing the
surface densities determined with pathfinder surveys to be scaled to
SKA-Mid assuming a Euclidean distribution of stellar radio sources.
The instantaneous surface density of M-dwarfs on \SI{1}{\hour}
time scales should therefore be at least \SI{0.4}{\deg^{-2}} in
Band 1 and \SI{1}{\deg^{-2}} in Band 2. Rarer high luminosity
M-dwarf bursts of order \SI{e17}{\erg\per\second\per\hertz} 
will be detectable out to $\sim$\SI{1}{\kilo\parsec}, though at this
distance source density is expected to drop with Galactic height and
thus surveys will be optimised at low Galactic latitude. 

These predicted rates are conservative due to incompleteness in the
ASKAP derived radio luminosity distribution below
$\sim$\SI{e15}{\erg\per\second\per\hertz} and the detection fraction
distribution below $\sim$\SI{0.07}, where detection fraction is the 
ratio of detections to observations of each star reported in \citet{2024MNRAS.tmp..161P}. This orders of magnitude increase
in the coverage of the radio star luminosity distribution will allow
improvements in classification of individual bursts and their energy
source, and repeat detections will allow the burst distributions of
individual stars to be established and tied to fundamental stellar
parameters (e.g. spectral type, age, and rotation) and
multi-wavelength activity indicators (e.g. optical flaring, starspot
coverage, and X-ray activity).

\begin{figure}
\includegraphics[width=0.8\columnwidth]{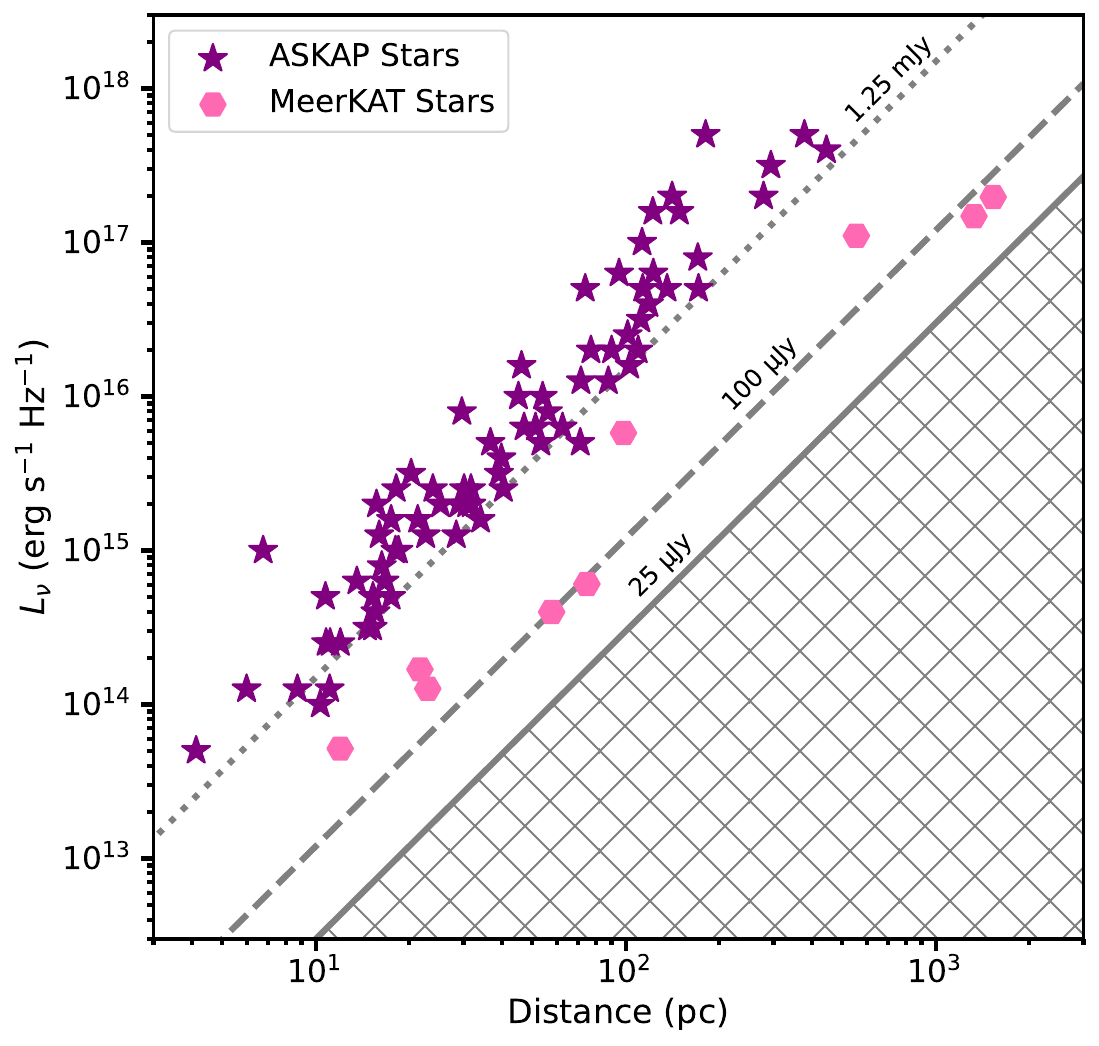}
\caption{The luminosities of radio stars detected by ASKAP \citep{2024MNRAS.tmp..161P} and a range of MeerKAT studies \citep{Andersson_Star,Driessen_Flareyboi, Driessen_EXO0408,Chastain2023giantstar,Chastain2025transients,Smirnov2025fastimaging}. The dotted, dashed and solid diagonal lines correspond respectively to typical $5\sigma$ sensitivity limits for ASKAP-VAST's 12~min images (1.25~mJy), MeerKAT-ThunderKAT's 15~min L-band images (100~$\mu$Jy) and predictions for the SKA-Mid \aastar\ for a 15~min image with Band 2 ($25~\mu$Jy). Most of these stars are variable and here their observed lower luminosities are plotted.
}
\label{fig:lum_vs_dist}
\end{figure}

We can similarly use the detections of stellar radio emission using ASKAP and MeerKAT to predict the sensitivity limits of transient surveys using 15 minute SKA Mid \aastar\ observations. Fig. \ref{fig:lum_vs_dist} presents stellar radio emission detections using ASKAP and MeerKAT in a luminosity-distance parameter space. This figure demonstrates the benefits of a `wedding cake' strategy, whereby wide-fields can survey many local stars whilst deeper observations allow us to probe further away and inspect a larger volume. 

We can also consider the sky density of detections by MeerKAT to consider the number of sources we may detect with SKA Mid \aastar. The approximate sky density of sources detected by MeerKAT in state-of-the-art surveys such as in \cite{Heywood2022} is $\sim6000$~sources deg$^{-2}$ when reaching $\sim1\mu$Jy sensitivity. The SKA-Mid Array \aastar\ is predicted to reach the same depth when observing with Band 2 in approximately 3\,h. If the array were to conduct a year-long survey that observed successive fields four times to detect variability, with each observation being 3\,h long, source counts would reach upwards of 4,000,000 sources (over less than 1/40th of the sky). If
time-variable phenomena make up a conservative 1\% of all sources
in a given radio image at any sensitivity limit \citep{Ofek2011, Driessen2022_variables}, this means that the
SKA-Mid will see up to $\sim40,000$ transients and variables in its
first year of operation. The majority of these will be scintillating AGN at $\sim40\%$, but the next most common variable found in image searches is stars at $\sim33\%$ \citep{2026arXiv260222739D}. This would result in 13,000 radio stars in such a survey.

\subsection{Low frequencies}

We can use stellar radio emission searches using LOFAR to estimate the detection rates using SKA-low \aastar. Note that searches using LOFAR have focused on circular polarisation searches (see Section\,\ref{sec: circ pol}). These predictions are therefore based on circular polarisation searches only.

We can model an SKA-low \aastar\ survey after the LOFAR Two-metre Sky Survey \citep[LoTSS;][]{2017A&A...598A.104S}.
If SKA-Low \aastar\ conducts a wide-field survey with 8\,h pointings at
144\,MHz with a bandwidth of $\approx$50 MHz, a 5$\sigma$ Stokes V detection
will need to be $\approx$50 $\mu$Jy based on current expected array sensitivity for \aastar\ and a robust parameter of 0. Therefore, we predict
that a southern sky (below declination 0$^{\circ}$) SKA-Low survey
could detect 2000$\pm$1000 circularly polarised sources, of which
at least 1000 are expected to be stellar in origin. We have factored the increase in the density of pulsars towards the Galactic
Plane into this prediction. 

Since SKA-Low is expected to become confusion-limited in total intensity in under an hour, a wide-field survey using 8~h pointings could be unlikely. Instead, if a wide-field survey is carried out with 1~h pointings at 144\,MHz and a bandwidth of $\approx 50$\,MHz, we anticipate about $700 \pm 300$ detections. This estimate assumes a $\geq 5\sigma$ Stokes~V detection threshold of $\geq 140~\mu\mathrm{Jy}$ and that the variable circularly polarised sources identified in 8\,h epochs are detected at the same rate in 1\,h epochs. Such a survey with repeat observations of the sky would detect more radio stars due to their variability.

\section{Summary}

Radio variability is a key property to consider when investigating and searching for radio stars. This variability also makes them one of the most commonly detected variable sources in untargeted, image-plane radio transient searches -- second only to AGN scintillation. It is therefore essential to consider multi-epoch observations and dynamic spectra when searching for and investigating radio stars. We predict that, between the SKA-mid and SKA-low \aastar\ arrays, surveys with a few-hours long observations and a modest number of repeat observations will detects thousands of radio stars. The sensitivity of the \aastar\ arrays will also result in detections of stars at more than double the distances achieved by current SKA precursor and pathfinder instruments and probe flares at luminosities more than an order of magnitude fainter than can currently be detected. These factors mean that untargeted searches for radio transients and variables and for radio stars specifically will result in unprecedented detections of radio stars. Identifying radio emission from a large population of stars is key to furthering the science goals of the stellar radio astronomy community; these are presented in detail in the chapters on ``Discovering and characterising exoplanets and ultracool dwarfs with the Square Kilometre Array'' \citep{Kavanagh01.2026.SKA}, ``Coherent and Incoherent Emission from the Ordered Magnetospheres of Low-Mass Stars, UCDs, and Massive Stars'' \citep{Cavallaro01.2026.SKA}, and ``Radio emission from star-planet interaction'' \citep{Vedantham01.2026.SKA}.

\section{Acknowledgements}

Part of this research was supported by the Australian Research Council Centre of Excellence for Gravitational Wave Discovery (OzGrav), project number CE230100016.

This scientific work uses data obtained from Inyarrimanha Ilgari Bundara, the CSIRO Murchison Radio-astronomy Observatory. We acknowledge the Wajarri Yamaji People as the Traditional Owners and native title holders of the Observatory site. CSIRO’s ASKAP radio telescope is part of the Australia Telescope National Facility (https://ror.org/05qajvd42). Operation of ASKAP is funded by the Australian Government with support from the National Collaborative Research Infrastructure Strategy. ASKAP uses the resources of the Pawsey Supercomputing Research Centre. Establishment of ASKAP, Inyarrimanha Ilgari Bundara, the CSIRO Murchison Radio-astronomy Observatory and the Pawsey Supercomputing Research Centre are initiatives of the Australian Government, with support from the Government of Western Australia and the Science and Industry Endowment Fund.

The MeerKAT telescope is operated by the South African Radio
Astronomy Observatory (SARAO), which is a facility of the National Research Foundation, an agency of the Department of Science and Innovation.

This paper is based (in part) on data obtained with the LOFAR telescope (LOFAR-ERIC). LOFAR \citep{2013A&A...556A...2V} is
the Low Frequency Array designed and constructed by ASTRON. It has observing, data processing, and data storage facilities in
several countries, that are owned by various parties (each with their own funding sources), and that are collectively operated by the
LOFAR European Research Infrastructure Consortium (LOFAR-ERIC) under a joint scientific policy. The LOFAR-ERIC resources
have benefited from the following recent major funding sources: CNRS-INSU, Observatoire de Paris and Université d’Orléans,
France; Istituto Nazionale di Astrofisica (INAF), Italy; BMBF, MIWF-NRW, MPG, Germany; Science Foundation Ireland (SFI),
Department of Business, Enterprise and Innovation (DBEI), Ireland; NWO, The Netherlands; The Science and Technology Facilities
Council, UK; Ministry of Science and Higher Education, Poland.

\bibliographystyle{abbrvnat-maxbibnames4}
\bibliography{radiostars} 

\end{document}